\documentclass[final,5p,times,twocolumn]{elsarticle}
 \biboptions{comma,sort&compress}
\usepackage{graphicx}
\usepackage{here}
\def\beq{\begin{equation}}
\def\eeq{\end{equation}}
\def\bea{\begin{eqnarray}}
\def\eea{\end{eqnarray}}
\def\bq{\begin{quote}}
\def\eq{\end{quote}}

\def\nnb{\nonumber}
\def\ga{\left(}
\def\dr{\right)}

\def\nnb{\nonumber}
\def\la{\langle}
\def\ra{\rangle}
\def\nin{\noindent}
\def\ba{\vspace*{-0.2cm}\begin{array}}
\def\ea{\end{array}\vspace*{-0.2cm}}

\def\b{$\bullet~$}
\def\als{\alpha_s}

\def\gg2{ \la\alpha_s G^2 \ra}
\def\gg3{g^3f_{abc}\la G^aG^bG^c \ra}
\def\ggg4{\la\als^2G^4\ra}

\def\beq{\begin{equation}}
\def\enq{\end{equation}}
\def\beqa{\begin{eqnarray}}
\def\enqa{\end{eqnarray}}
\def\nnb{\nonumber}

\def\lb{\label}

\newcommand{\rag}{\rangle}
\newcommand{\lag}{\langle}

\journal{Elsevier}

\begin{document}

\begin{frontmatter}

\title{Summary on $\overline{m}_{c,b}(\overline{m}_{c,b})$ and precise $f_{D_{(s)},B_{(s)}}$  from  heavy-light QCD spectral sum rules 
$^*$ \corref{cor0}
}
\cortext[cor0]{Talk given at the 16th International Conference in QCD (QCD12), 2-6th july 2012 (Montepllier-FR). This talk  is a summary of the results in \cite{SNFB12} where more detailed discussions and more complete references can be found.}
 \author[label2]{Stephan Narison}
\address[label2]{Laboratoire
Particules et Univers de Montpellier, CNRS-IN2P3, 
Case 070, Place Eug\`ene
Bataillon, 34095 - Montpellier, France.}
   \ead{snarison@yahoo.fr}
\begin{abstract}
\nin
We summarize recent results obtained in \cite{SNFB12} on the running $\overline{m}_{c,b}(\overline{m}_{c,b})$ in the $\overline{MS}$ scheme and $f_{D_{(s)}, B_{(s)}}$ using QCD spectral sum rules (QSSR) known to N2LO PT series, including all dimension-six NP condensate contributions in the full QCD theory, an estimate of the N3LO terms based on the geomteric growth of the PT series  and using the most recent values of the QCD input parameters given in Table \ref{tab:param}. 
The study of the effects of the subtraction scale $\mu$  on ``different QSSR data" and the use (for the first time) of the Renormalization Group Invariant (RGI) scale independent quark masses in the analysis
are emphasized. The estimates [rigourous model-independent upper bounds] reported in Table~\ref{tab:result}: $f_D/f_\pi=1.56(5)[\leq 1.68(1)]$,  $f_B/f_\pi=1.58(5)[\leq 1.80(3)]$ and $f_{D_s}/f_K= 1.58(4) [\leq 1.63(1)]$, 
$f_{B_s}/f_K=1.50(3)[\leq 1.61(3.5)]$, improve previous QSSR estimates. The remarkable agreements with some of the experimental data on $f_{D,D_s}$ and  with lattice simulations within dynamical quarks confirm both the success of the QSSR semi-approximate approach based on the OPE in terms of the quark and gluon condensates and of the Minimal Duality Ansatz (MDA) for parametrizing the hadronic spectral function which we have tested from the complete data of the $J/\psi$ and $\Upsilon$ systems.  The running quark masses $\overline{m}_c({m_c})=1286(66) $ MeV and  $\overline{m}_b({m_b})= 4236(69)$ MeV from  $M_{D,B}$ are in good agreement though less accurate than the ones from recent $J/\psi$ and $\Upsilon$ sum rules. 
\end{abstract}
\begin{keyword}  
QCD spectral sum rules, meson decay constants, heavy quark masses. 
\end{keyword}
\end{frontmatter}
\section{Introduction and a short historical review}
The (pseudo)scalar meson decay constants $f_P$ are of prime interests for understanding the realizations of chiral symmetry in QCD. 
In addition to the well-known values of $f_\pi$ and $f_K$ which control  the light flavour chiral symmetries, it is also desirable to extract the ones of the heavy-light charm and bottom quark systems with high-accuracy. These decay constants are normalized through the matrix element:
\beq
\la 0|J^P_{\bar qQ}(x)|P\ra= f_P M_P^2~:~J^P_{\bar qQ}(x)\equiv (m_q+M_Q)\bar q(i\gamma_5)Q~,
\label{eq:fp}
\label{eq:current}
\eeq
where: 
$
J^P_{\bar qQ}(x)
$
is the local heavy-light pseudoscalar current;  $q\equiv d,s;~Q\equiv c,b;~P\equiv D_{(s)}, B_{(s)}$ and where $f_P$  is related to  the leptonic width $\Gamma (P^+\to l^+\nu_l)$.
Since the original works of 
NSV2Z \cite{NSV2Z} based on QCD spectral sum rules (QSSR) \cite{SVZ}\,\footnote{For reviews, see e.g: \cite{RRY,SNB1,SNB2,SNB3}.}, a large amount of QSSR works have been devoted to give bounds on $f_P$ \cite{SNZ, BROAD,GENERALIS} and to estimate their values \cite{SNFB,SNFB2,SNFB4}. More recent works including N2LO PT corrections have been derived later on in full QCD \cite{SNFB3,JAMIN3,KHOJ} and in HQET~\cite{PENIN}. 
Different earlier papers \cite{SNB1,SNB2} have been scrutinized  in \cite{SNFB,SNB2}, where Narison found that the apparent discrepancies between the different QSSR results can be solved if one applies carefully the stability  criteria (also called sum rule window) of the results versus the external QSSR variables and continuum threshold $t_c$. In this way, and for given values of $m_{c,b}$, he obtained the values:
$ f_D\simeq (1.31\pm 0.12)f_\pi~,~ f_B\simeq (1.6\pm 0.1)f_\pi~,$
which are independent of the forms of the sum rules used. However, the result has been quite surprising as it indicates a large violation of the heavy quark symmetry scaling predictions \cite{HQET}, where $1/M_Q$ corrections have been estimated numerically in \cite{SNFB4}.  This ``unexpected result" has been confirmed few years later by lattice calculations \cite{MARTI}. 
In this talk, we present improved estimates and bounds on $f_P$ and on $\overline{m}_{c,b}({m_{c,b}})$ obtained recently in \cite{SNFB12} in full QCD theory, where
 the most recent values of the (non-)perturbative QCD parameters given in Table \ref{tab:param}
 have been used.  The expressions of NLO PT in \cite{BROAD,BROAD1}, of N2LO PT  in  \cite{CHET}, of the non-perturbative in \cite{NSV2Z,GENERALIS} and the light quark mass corrections in \cite{BROAD,GENERALIS,JAMIN} have been used. The N3LO PT contributions have been estimated by assuming the  geometric growth of the series \cite{NZ} which is dual to the effect of a $1/q^2$ term \cite{CNZ,ZAK}.  Like in \cite{SNFB2}, the previous PT results obtained in the on-shell scheme is translated to the one in the $\overline{MS}$-scheme by using the relation between the on-shell and $\overline{MS}$ mass known at present to N3LO \cite{SNB1,SNB2}. 
The Renormalization Group Invariant (RGI) $s,c,b$ quark masses introduced by \cite{FNR} which are scale and (massless) scheme independent have been also used for the first time, while a careful study of the effect of the substraction scale on each ``QSSR data point" has been performed. 
\vspace*{-0.5cm}
\section{QCD spectral  sum rules (QSSR)}
\vspace*{-0.2cm}
\subsection*{\b Forms of the  sum rules}
\nin
We shall be concerned with the two-point correlator :
\beq
\psi^{P}_{\bar qQ}(q^2)=i\int d^4x ~e^{iq.x}\lag 0
|TJ^P_{\bar qQ}(x)J^P_{\bar qQ}(0)^\dagger
|0\rag~,
\lb{2po}
\eeq
where $J_{\bar qQ}(x)$ is the local current defined in Eq. (\ref{eq:current}). 
The associated Laplace sum rules (LSR)  ${\cal L}_{\bar qQ}(\tau)$ and
its ratio ${\cal R}_{\bar qQ}(\tau)$ read\,\cite{SVZ}\,\footnote{Radiative corrections to the exponential sum rules have been first derived in \cite{SNRAF}, where it has been noticed that the PT series has the property of an Inverse Laplace transform.}:
\bea
{\cal L}_{\bar qQ}(\tau,\mu)&=&\int_{(m_q+M_Q)^2}^{t_c}dt~e^{-t\tau}\frac{1}{\pi} \mbox{Im}\psi^P_{\bar qQ}(t,\mu)~,\nnb\\
{\cal R}_{\bar qQ} (\tau,\mu) &=& \frac{\int_{(m_q+M_Q)^2}^{t_c} dt~t~ e^{-t\tau}\frac{1}{\pi}\mbox{Im}\psi^P_{\bar qQ}(t,\mu)}
{\int_{(m_q+M_Q)^2}^{t_c} dt~ e^{-t\tau} \frac{1}{\pi} \mbox{Im}\psi_{\bar qQ}(t,\mu)}~,
\label{eq:lsr}\label{eq:ratiolsr}
\eea
where $\mu$ is the subtraction point which appears in the approximate QCD series when radiative corrections are included. 
 The ratio of sum  rules ${\cal R}_{\bar qQ} (\tau,\mu)$ is useful, as it is equal to the
resonance mass squared, in
  the Minimal Duality Ansatz (MDA) parametrization of the spectral function:
\beq
\frac{1}{\pi}\mbox{ Im}\psi^P_{\bar qQ}(t)\simeq f^2_PM_P^4\delta(t-M^2_P)
  \ + \
  ``\mbox{QCD cont.}" \theta (t-t_c),
\label{eq:duality}
\eeq
where $f_P$ is the decay constant defined in Eq. (\ref{eq:fp}) and the higher states contributions are smeared by the ``QCD continuum" coming from the discontinuity of the QCD diagrams and starting from a constant threshold $t_c$. 
We shall also use for the $B$-meson, the moments obtained after deriving $n+1$-times the two-point function and evaluated at $Q^2=0$ (MSR) \cite{SVZ}, where an expansion in terms of the on-shell mass $M_b$ can be used. They read:
\bea
{\cal M}^{(n)}_{\bar qb}(\mu)&=&\int_{(m_q+M_b)^2}^{t_c}{dt\over t^{n+2}}~\frac{1}{\pi} \mbox{Im}\psi^P_{\bar qb}(t,\mu)~,\nnb\\
{\cal R}^{(n)}_{\bar qb}(\mu)&=&{\int_{(m_q+M_b)^2}^{t_c}{dt\over t^{n+2}}~\frac{1}{\pi} \mbox{Im}\psi^P_{\bar qb}(t,\mu)\over 
\int_{(m_q+M_b)^2}^{t_c}{dt\over t^{n+3}}~\frac{1}{\pi} \mbox{Im}\psi^P_{\bar qb}(t,\mu)}~.
\label{eq:mom}
\label{eq:momratio}
\eea
\vspace*{-0.2cm}
\subsection*{\b Test of the Minimal Duality Ansatz (MDA) from $J/\psi$ and $\Upsilon$}\label{sec:duality}
\begin{figure}[hbt] 
\begin{center}
\centerline {\hspace*{-7.5cm} a) }\vspace{-0.3cm}
{\includegraphics[width=4.4cm  ]{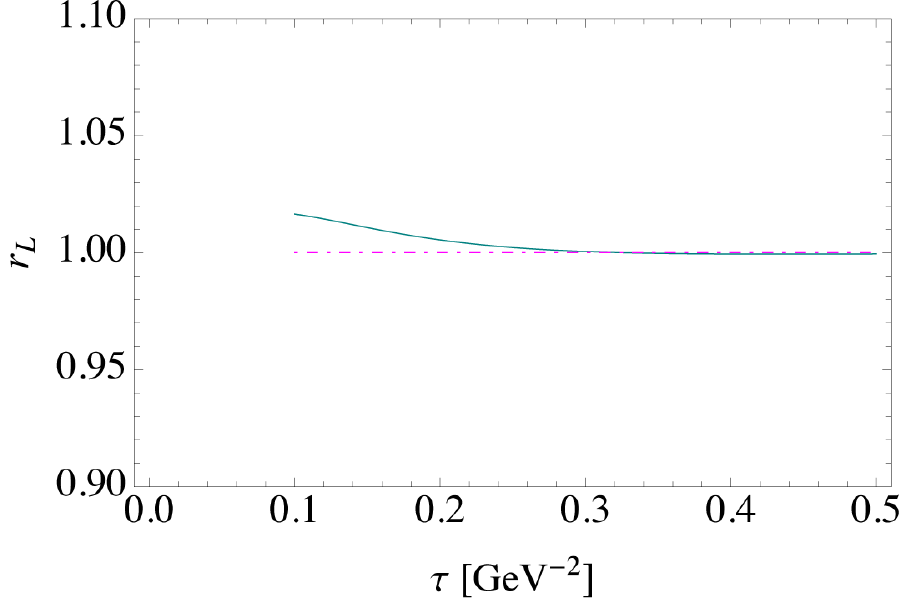}}
\centerline {\hspace*{-7.5cm} b) }\vspace{-0.3cm}
{\includegraphics[width=4.4cm  ]{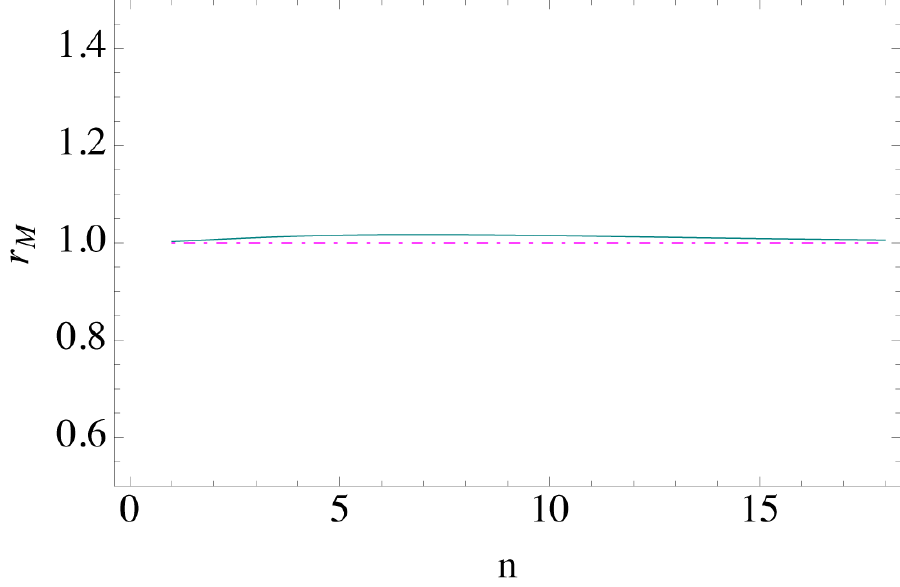}}
\caption{
\scriptsize 
{\bf a)} $\tau$-behaviour of the ratio of  ${\cal L}^{exp}_{\bar bb}/ {\cal L}^{dual}_{\bar bb}$ for $\sqrt{t_c}= M_{\Upsilon(2S)}$-0.15 GeV. The red dashed curve corresponds to the strict equality for all values of $\tau$.; {\bf b)} the same as a) but for ${\cal M}^{(n)exp}_{\bar bb}/ {\cal M}^{(n)dual}_{\bar bb}$ versus the number of derivatives $n$.
}
\label{fig:bduality} 
\end{center}
\end{figure}

\nin
The MDA presented in Eq. (\ref{eq:duality}), when applied to the $\rho$-meson reproduces within 15\% accuracy the ratio ${\cal R}_{\bar dd}$ measured from the total cross-section $e^+e^-\to {\rm I=1 ~hadrons}$ data (Fig. 5.6 of \cite{SNB2}), while in the case of charmonium, $M_\psi^2$ from ${\cal R}^{(n)}_{\bar cc}$ has been compared with the one from complete data where a remarkable agreement for higher $n\geq 4$ values (Fig. 9.1 of \cite{SNB2}) has been found. Tests of MDA from the $J/\psi$ and $\Upsilon$ systems have been also done in \cite{SNFB12}. Taking $\sqrt{t_c}\simeq M_{\Upsilon(2S)}- 0.15$ GeV, we show (for instance)  the ratio between ${\cal L}^{exp}_{\bar bb}$ and  ${\cal L}_{\bar bb}^{dual}$ in Fig. (\ref{fig:bduality}a) for LSR and ${\cal M}_{\bar bb}^{(n)exp}$ and  ${\cal M}_{\bar bb}^{(n)dual}$ for MSR in Fig. (\ref{fig:bduality}b) for the $\Upsilon$ systems indicating that for heavy quark systems the r\^ole of the QCD continuum will be smaller than in the case of light quarks and the exponential weight or high number of derivatives suppresses efficiently the QCD continuum contribution but enhances the one of the lowest ground state in the spectral integral. 
 We have used the simplest QCD continuum 
 expression for massless quarks to order $\alpha_s^3$ from the threshold $t_c$ \cite{SNH10}\,\footnote{We have checked that the spectral function including complete mass corrections give the same results.}:
$
 {\rm QCD~ cont.}=1+as+1.5as^2-12.07as^3.
$
 One can see that the MDA, with a value of $\sqrt{t_c}$ around the value of the 1st radial excitation mass, describes quite well the complete data in the region of $\tau$ and $n$ where the corresponding sum rules present $\tau$ (in units of (${\rm GeV}^{-2}$) or $n$ stability\,\cite{SNH10}:
\beq
  \tau^\psi\simeq (1.3\sim 1.4)~,~
  ~ \tau^\Upsilon\simeq (0.2\sim 0.4)~,~~n^\Upsilon\simeq (5\sim 7),
  \label{eq:tau}
 \eeq
  as we shall see later on. 
 Moreover, MDA  has been also used in \cite{PERIS} (called Minimal Hadronic Ansatz in this paper) in the context of large $N_c$ QCD, where 
 it provides a very good approximation to the observables one compute. 
\nin
\vspace*{-0.5cm}
\begin{figure}[hbt] 
\begin{center}
{\includegraphics[height=5cm  ]{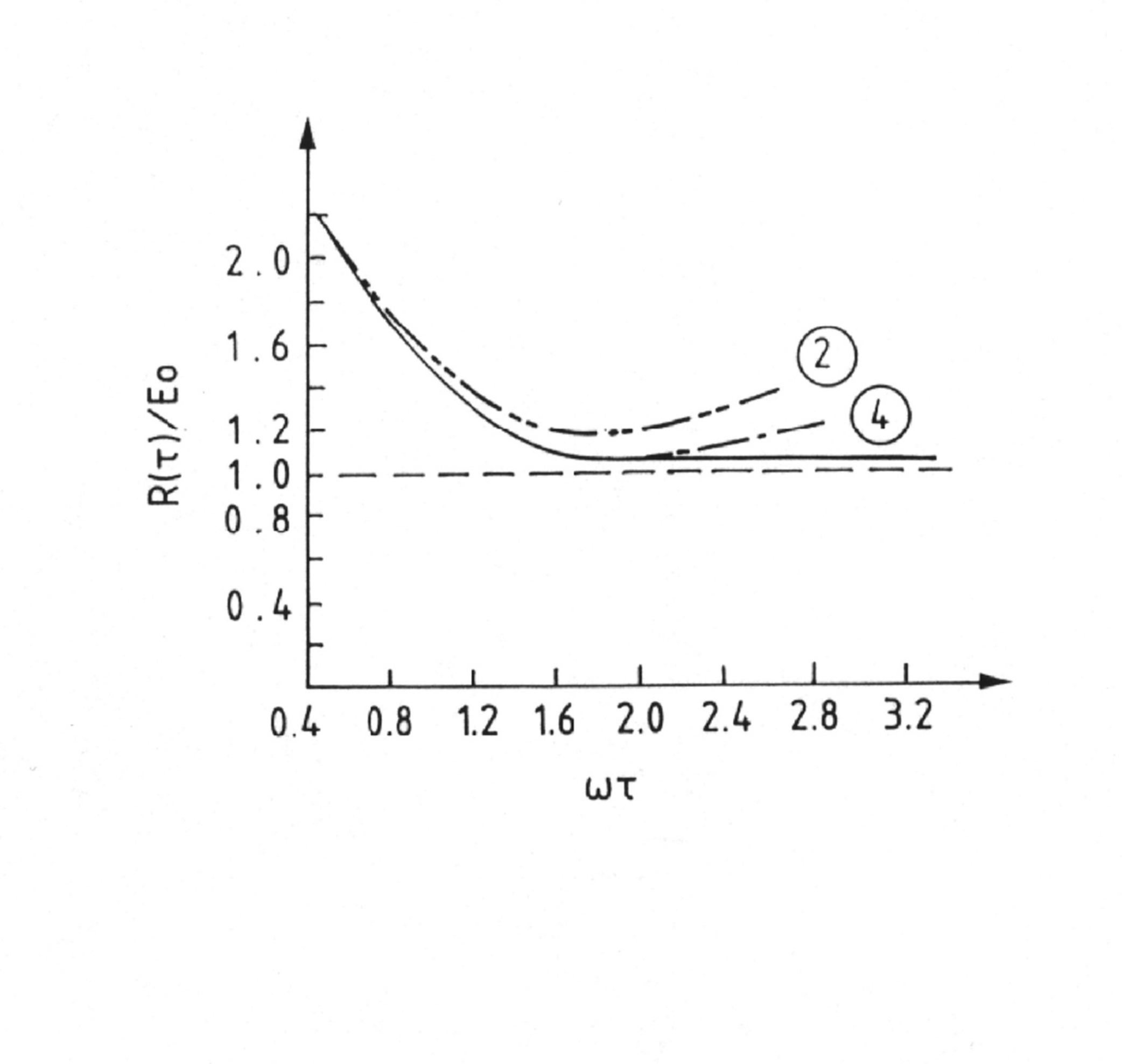}}
\vspace*{-1.5cm}
\caption{
\scriptsize 
{\bf a)} $\tau$-behaviour of ${\cal R}(\tau)$ normalized to the ground state energy $E_0$ for the harmonic oscillator. 2 and 4 indicate the number of terms in the approximate series.}
\label{fig:oscillo} 
\end{center}
\end{figure} 
\nin
\vspace*{-0.7cm}
\subsection*{\b Optimal results from stability criteria}
\nin
Using the theoretical expressions of ${\cal L}_{\bar dQ}^{th}$ or ${\cal M}_{\bar db}^{(n)th}$, and parametrizing its experimental side ${\cal L}_{\bar dQ}^{exp}$ or ${\cal M}_{\bar db}^{(n)exp}$ by the MDA in  Eq. (\ref{eq:duality}), one can extract the decay constant $f_P$ and the RGI quark mass $\hat m_Q$.
In principle the equality  ${\cal L}_{\bar dQ}^{th}={\cal L}_{\bar dQ}^{exp}$ should be satisfied for any values of the external (unphysical) set of variables $(\tau, t_c)$, if one knows exactly ${\cal L}_{\bar dQ}^{th}$ and ${\cal L}_{\bar dQ}^{exp}$. Here, we make an analogy with the harmonic oscillator, where  the ratio of moments ${\cal R}_{\bar dQ}$ is a function of the imaginary time variable $\tau$ and where one knows the exact and approximate results. One can find \cite{BELL} that the exact energy $E_0$ of the ground state can be approached from above by the approximate series
(see Fig. \ref{fig:oscillo}).  At the minimum or inflexion point (stability) of the curves, one has a ground state dominance. For small time (large $Q^2$), all level contributes, while for large time (small $Q^2$) the series breakdown.
We shall apply this stability criterion inspired from quantum mechanics in our analysis. 
In principle, the continuum threshold $\sqrt{t_c}$ in  Eq. (\ref{eq:duality}) is a free parameter, though one expects its value to be around the mass of the 1st radial excitation because the QCD spectral function is supposed to smear all the higher state contributions in the spectral integral as explicitly shown previously in Section \ref{sec:duality}. In order to avoid the model-dependence on the results, Refs. \cite{SNFB,SNFB2,SNFB3,SNFB4,SNB1,SNB2} have considered the conservative range of $t_c$-values where one starts to have $\tau$- or $n$-stability until which one reaches a $t_c$-stability where the contribution of the lowest ground state to the spectral integral completely dominates. For the $D$ and $B$ mesons, this range is: 
\beq
 t_c^D\simeq (5.5 \sim 9.5)~{\rm GeV}^2,~~~~~t_c^B\simeq (33\sim 45) ~{\rm GeV}^2.
 \label{eq:tc}
\eeq
\vspace*{-0.5cm}
\section{The QCD input parameters}
\vspace*{-0.2cm}
We shall use the QCD parameters (with obvious notations) and their values given in Table {\ref{tab:param}}, where  $\rho\simeq 2$ indicates the deviation from the four-quark vacuum saturation.  From the running ${\bar m}_{q,Q}$ quark parameters, one can deduce the corresponding RGI quantities ${\hat m}_{q,Q}$ and $\hat\mu_q$ \cite{FNR} known to order $\alpha_s^3$ \cite{SNB1,SNB2,RUNDEC}:
$
{\bar m}_{q,Q}(\tau)=
{\hat m}_{q,Q}  \ga-\beta_1a_s\dr^{-2/{
\beta_1}}\times C(a_s),~
{\la\bar qq\ra}(\tau)=-{\hat \mu_q^3  \ga-\beta_1a_s\dr^{2/{
\beta_1}}}/C(a_s),~
{\la\bar qg\sigma Gq\ra}(\tau)=-{M_0^2{\hat \mu_q^3} \ga-\beta_1a_s\dr^{1/{3\beta_1}}}/C(a_s)$~,
where $\beta_1=-(1/2)(11-2n_f/3)$ is the first coefficient of the $\beta$ function 
for $n_f$ flavours; $a_s\equiv \alpha_s(\tau)/\pi$. The QCD correction factor $C(a_s)$ is $1+1.1755a_s+1.5008a_s^2 +...$, for $n_f=5$ flavours and shows a good convergence. Therefore,
the RGI quantities  to order $\alpha_s^2$ (heavy quarks) and to order $\alpha_s$ (light quarks), in units of MeV are:
$
\hat m_c=1467(14),~\hat m_b=7292(14),~
\hat m_s=128(7),~\hat\mu_q=251(6)~.
$
\vspace*{-0.5cm}
{\scriptsize
\begin{table}[hbt]
\setlength{\tabcolsep}{0.2pc}
 \caption{
QCD input parameters. 
 }
    {\small
\begin{tabular}{lll}
&\\
\hline
Parameters&Values& Ref.    \\
\hline
$\alpha_s(M_\tau)$& $0.325(8)$&\cite{SNTAU,BNP,BETHKE}\\
$\overline{m}_s(2)$&$96.1(4.8)$ MeV&average \cite{SNmass}\\
$\overline{m}_c(m_c)$&$1261(12)$ MeV &average \cite{SNH10}\\
$\overline{m}_b(m_b)$&$4177(11)$ MeV&average \cite{SNH10}\\
${1\over 2}\la \bar uu+\bar dd\ra^{1/3} (2)$&$-(275.7\pm 6.6)$ MeV&\cite{SNB1,SNmass}\\
$\la\bar ss\ra/\la\bar dd\ra$&0.74(3)&\cite{SNB1,SNmass,Hbaryon}\\
$M_0^2$&$(0.8 \pm 0.2)$ GeV$^2$&\cite{JAMI2,HEID,SNhl}\\
$\la\alpha_s G^2\ra$& $(7\pm 1)\times 10^{-2}$ GeV$^4$&
\cite{SNTAU,LNT,SNH10,SNI,BELL}\\
$\la g^3  G^3\ra$& $(8.2\pm 1.0)$ GeV$^2\times\la\alpha_s G^2\ra$&
\cite{SNH10}\\
$\rho \la \bar qq\ra^2$&$(4.5\pm 0.3)\times 10^{-4}$ GeV$^6$&\cite{SNTAU,LNT,JAMI2}\\
\hline
\end{tabular}
}
\label{tab:param}
\end{table}
}
\vspace*{-0.5cm}
\begin{figure}[hbt] 
\begin{center}
\centerline {\hspace*{-7.5cm} a) }\vspace{-0.3cm}
{\includegraphics[width=4.8cm  ]{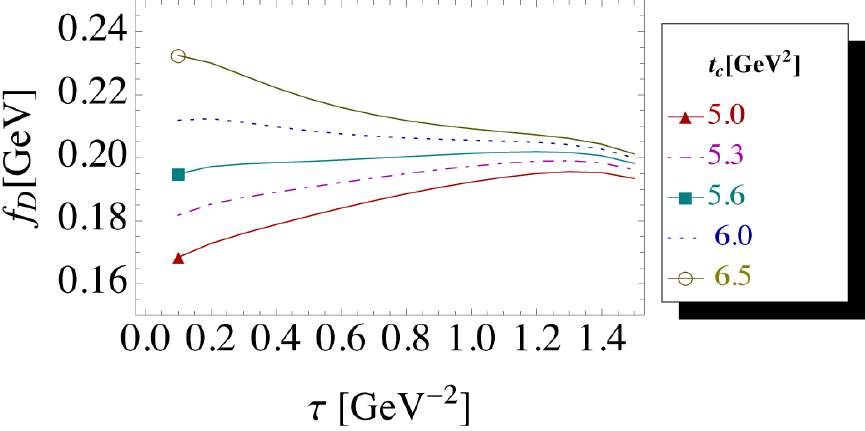}}
\centerline {\hspace*{-7.5cm} b) }\vspace{-0.3cm}
{\includegraphics[width=4.8cm  ]{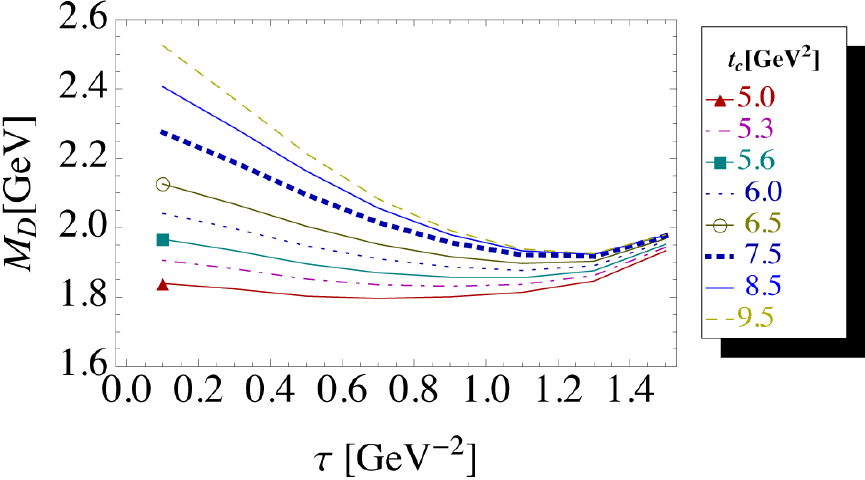}}
\caption{
\scriptsize 
{\bf a)} $\tau$-behaviour of $f_D$ from  ${\cal L}_{\bar dc} $ for different values of $t_c$, for a given value of the subtraction point $\mu=\tau^{-1/2}$ and for $\hat m_c=1467$ MeV; {\bf b)} the same as a) but for $M_D$ from ${\cal R}_{\bar dc}$. }
\label{fig:fdtau} 
\end{center}
\end{figure} 
\nin
\begin{figure}[hbt] 
\begin{center}
\centerline {\hspace*{-7.5cm} a) }\vspace{-0.6cm}
{\includegraphics[width=4.8cm  ]{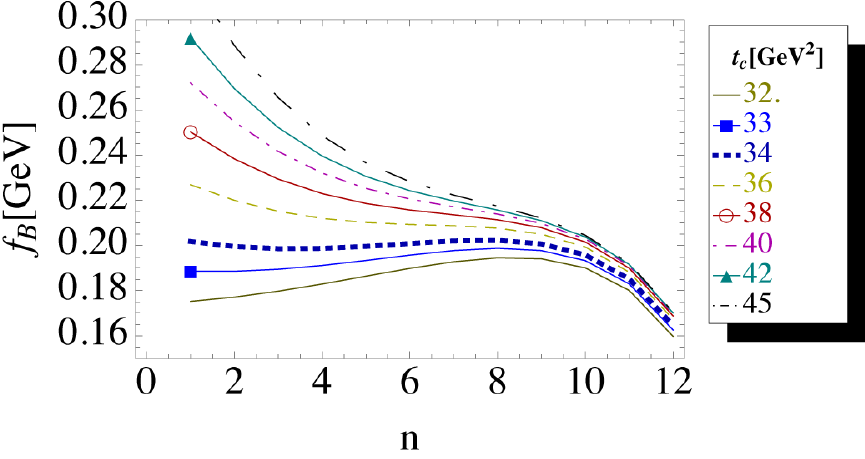}}
\centerline {\hspace*{-7.5cm} b) }\vspace{-0.3cm}
{\includegraphics[width=4.8cm  ]{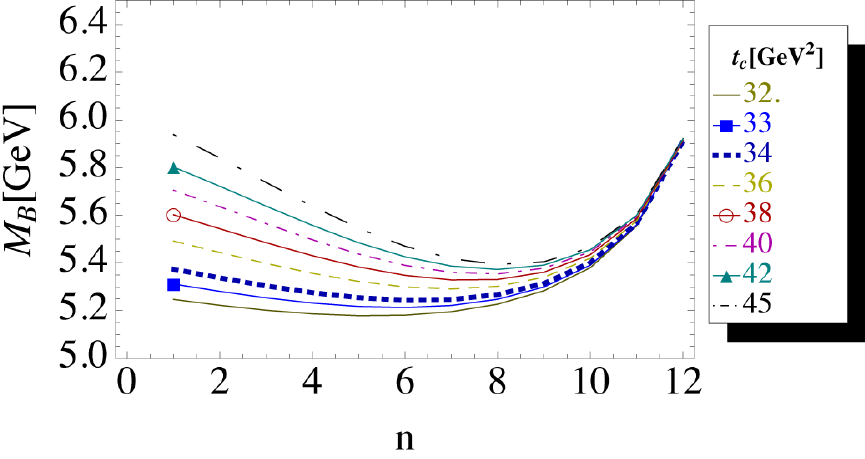}}
\caption{
\scriptsize 
{\bf a)} $n$ behaviour  of  $f_B$ from  MSR  for different values of $t_c$, for $\hat m_b$=7292 MeV and at the subtraction point $\mu=4$ GeV; {\bf b)} the same as a) but for $M_B$. }
\label{fig:fb_n} 
\end{center}
\end{figure} 
\nin
\section{QSSR analysis and results}
\begin{figure}[hbt] 
\begin{center}
\centerline {\hspace*{-7.5cm} a) }\vspace{-0.6cm}
{\includegraphics[width=4.4cm  ]{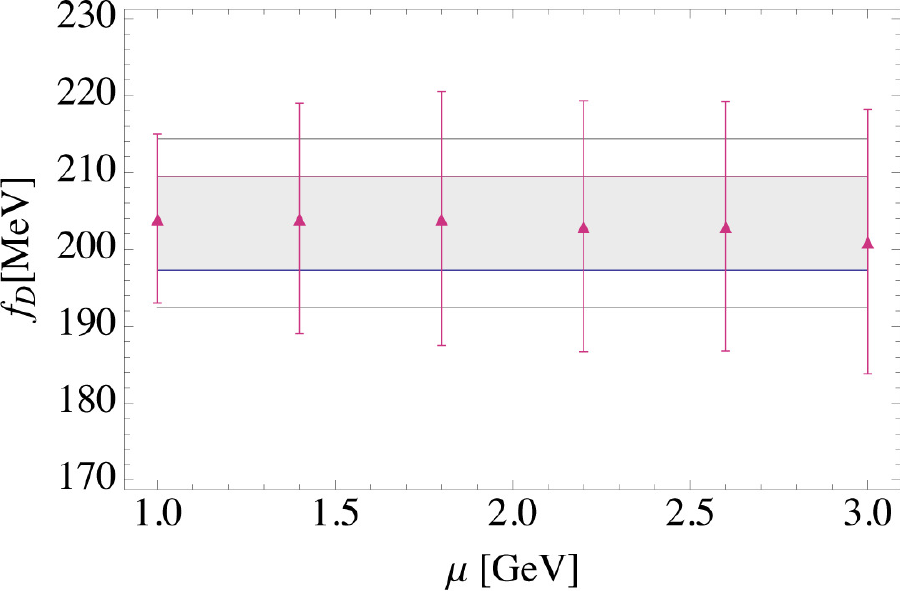}}
\centerline {\hspace*{-7.5cm} b) }\vspace{-0.3cm}
{\includegraphics[width=4.4cm  ]{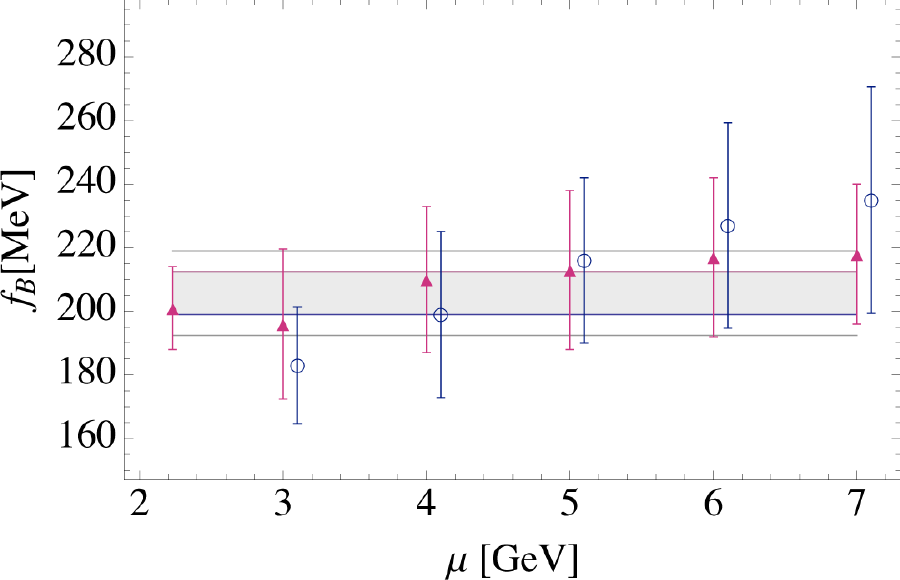}}
\caption{
\scriptsize 
{\bf a)} Values of  $f_D$ from LSR at different values of the subtraction point $\mu$ and for $\hat m_c$=1467 MeV:
{\bf b} Values of  $f_B$ from LSR (red triangle) and from MSR (blue open circle) at different values of the subtraction point $\mu$ and for $\hat m_b$=7292 MeV. }
\label{fig:fb_mean} 
\label{fig:fd_mean} 
\end{center}
\end{figure}

The QCD expressions of the sum rules are given in \cite{SNFB12} to order $\alpha_s^2$.
We illustrate the analysis for the LSR in Fig. \ref{fig:fdtau} for the $D$-meson and for the MSR 
in Fig.\ref{fig:fb_n} for the $B$-meson. The optimal values of $f_{D_{(s)}}$ and $f_{B_{(s)}}$ are obtained in the range of $\tau$ and $t_c$ values given in Eqs. (\ref{eq:tau}) and (\ref{eq:tc}). We study the dependence of the results on the values of the subtraction scale $\mu$ in Fig. (\ref{fig:fd_mean}).
\nin
\section{Summary and conclusions}
\vspace*{-0.5cm}
{\scriptsize
\begin{table}[H]
\setlength{\tabcolsep}{0.15pc}
 \caption{
Results from the open charm and beauty systems in units of MeV and comparison with experimental data and lattice simulations
using $n_f=2$ \cite{ETM,HEITGER} and $n_f=3$ \cite{HPQCD,MILC} dynamical quarks. $f_P$ are normalized as $f_\pi=130.4$ MeV and $f_K=156.1(9)$ MeV  \cite{ROSNER}. }
    {\small
\begin{tabular}{llll}
&\\
\hline
Charm&Bottom& Ref.    \\
\hline
$\overline{m}_c(\overline{m}_c)$&$\overline{m}_b(\overline{m}_b)$\\
1286(66) &4236(69)& This work \\
1280(40)&4290(140)&ETMC\cite{ETM}\\
$f_D$&$f_B$\\
$204(6)\equiv 1.56(5)f_\pi$&$206(7)\equiv 1.58(5)f_\pi$&This work\\
$\leq 218.4(1.4)\equiv 1.68(1)f_\pi$&$\leq 235.3(3.8)\equiv 1.80(3)f_\pi$& This work\\
207(9)&--& Data \cite{ROSNER,ASNER}\\
212(8)&195(12)& ETMC\cite{ETM}\\
--& 193(10)& ALPHA \cite{HEITGER} \\
207(4)&190(13)& HPQCD \cite{HPQCD}\\
219(11)&197(9)& FNAL \cite{MILC}\\
 $f_{D_s}$&$f_{B_s}$\\
$246(6)\equiv 1.59(5)f_K$&$234(5)\equiv 1.51(4)f_\pi$&This work\\
$\leq 253.7(1.5)\equiv 1.61(1)f_K$&$\leq 251.3(5.5)\equiv 1.61(4)f_K$& This work\\
260(5.4)&--& Data \cite{ROSNER,ASNER}\\
248(6)&232(12)& ETMC\cite{ETM}\\
--& 219(12)& ALPHA \cite{HEITGER}\\
248(2.5)&225(4)& HPQCD \cite{HPQCD}\\
260(11)&242(10)& FNAL \cite{MILC}\\

\hline
\end{tabular}
}
\label{tab:result}
\end{table}
}
\vspace*{-0.7cm}
We have re-extracted the decay constants $f_{D,D_{s}}$ and $f_{B,B_{s}}$ and the running
quark masses $\overline{m}_{c,b}(\overline{m}_{c,b})$ using QCD
spectral sum rules (QSSR), the recent values of the QCD (non-)perturbative  parameters given in Table \ref{tab:param} and (for the first time) the scale independent Renormalization Group Invariant (RGI) heavy quark masses  in the analysis after translating the on-shell PT expression of the spectral function to the $\overline{MS}$ scheme. 
We have noticed that $f_P$ are very sensitive to $m_Q$ and decreases when $m_Q$ increases. We have taken the conservative range of $t_c$ in Eq. (\ref{eq:tc}) covering the beginning of the $\tau$- or $n$-stability until the  $t_c$-stability  [Figs. (\ref{fig:fdtau}) and (\ref{fig:fb_n})]. We have carefully studied  the effects of the subtraction scale $\mu$ on the ``QSSR data" [Fig. (\ref{fig:fb_mean})]. Our final results  in Table \ref{tab:result}  agree and improve existing QSSR results in the literature. Large mass corrections responsible of $f_D\simeq f_B$ have been estimated in \cite{SNFB4}. $f_D$ and $f_{D_s}$ agree within the errors with the data compiled in \cite{ROSNER,ASNER}, while the upper bound on $f_{D_s}$  can already exclude some existing data and theoretical estimates (see e.g. \cite{SNFB8} for some attempts to extract $\vert V_{cd}\vert$ and $\vert V_{cs}\vert$). 
Our results are compatible (in values and precisions) with lattice simulations including dynamical quarks \cite{ETM,HEITGER,HPQCD,MILC}, which are not surprising as both methods study the same  two-point correlator though evaluated in two different space-times and
use the 1st principles of QCD (here is the OPE in terms of the quarks and gluon condensates which semi-approximate non-perturbative confinement). These agreements also confirm the accuracy of the MDA for describing the spectral function in the absence of complete data, which has been tested earlier \cite{SNB1,SNB2} and in this paper from the charmonium and bottomium systems and  in \cite{PERIS}  in the large $N_c$ limit of QCD . 

\end{document}